\documentclass[10pt,twocolumn,english,aps,prd,nofootinbib,preprintnumbers]{revtex4}
\usepackage[latin9]{inputenc}
\usepackage{graphicx}
\usepackage{amsmath}
\usepackage{amssymb}
\usepackage{amsfonts}
\usepackage{lscape}
\usepackage{hyperref}
\usepackage{url}
\usepackage{epsfig}
\usepackage{color}
\usepackage{bm}
\usepackage{tabularx}
\usepackage{mathtools}
\newcommand{\be}{\begin{equation}}
\newcommand{\ee}{\end{equation}}
\newcommand{\ba}{\begin{eqnarray}}
\newcommand{\ea}{\end{eqnarray}}
\newcommand{\nn}{\nonumber}

\newcommand{\Msun}{M_\odot}

\renewcommand{\[}{\begin{equation}}
\renewcommand{\]}{\end{equation}}

\makeatother

\begin{document}

\preprint{IFT-UAM/CSIC-17-115, CERN-TH-2017-250}

\title{Gravitational wave energy emission and detection rates of Primordial Black Hole hyperbolic encounters}

\author{Juan Garc\'ia-Bellido$^{(a,b)}$}
\email{juan.garciabellido@cern.ch}

\author{Savvas Nesseris$^{(a)}$}
\email{savvas.nesseris@csic.es}

\affiliation{$^{(a)}$Instituto de F\'isica Te\'orica UAM-CSIC, Universidad Auton\'oma de Madrid,
Cantoblanco, 28049 Madrid, Spain\\
$^{(b)}$CERN, Theoretical Physics Department, 1211 Geneva, Switzerland}

\date{\today}

\begin{abstract}
We describe in detail gravitational wave bursts from Primordial Black Hole (PBH) hyperbolic encounters. The bursts are one-time events, with the bulk of the released energy happening during the closest approach, which can be emitted in frequencies that could be within the range of both LIGO (10-1000Hz) and LISA ($10^{-6}-1$ Hz). Furthermore, we correct the results for the power spectrum of hyperbolic encounters found in the literature and present new exact and approximate expressions for the peak frequency of the emission. Note that these GW bursts from hyperbolic encounters between PBH are complementary to the GW emission from the bounded orbits of BHB mergers detected by LIGO, and help breaking degeneracies in the determination of the PBH mass, spin and spatial distributions.
\end{abstract}
\maketitle

\section{Introduction}

In the last couple of years, Advanced LIGO has brought forward a new era of Gravitational Wave Astronomy, with the observation of several massive Black Hole (BH) merger events~\cite{Abbott:2016blz,Abbott:2016nmj,TheLIGOScientific:2016pea,Abbott:2017vtc,Abbott:2017oio}, as well as the recent detection of a binary neutron star inspiral \cite{TheLIGOScientific:2017qsa}, which opened the era of Multimessenger Astronomy~\cite{GBM:2017lvd}. In the case of the BH mergers, the signal corresponds to the inspiralling of two massive BHs in approximately circular orbits, and the emission of gravitational waves (GW) leading to the final merger is in excellent agreement with the expected result from General Relativity (GR).

Since the large number of massive BH binaries were unexpected, see however~\cite{Belczynski:2009xy}, they seem to hint at a new population of massive BHs. This opens fertile ground for speculating that AdvLIGO could have observed Primordial Black Holes (PBH) as a significant fraction of Cold Dark Matter (CDM)~\cite{Bird:2016dcv,Clesse:2016vqa,Sasaki:2016jop}, thus offering a viable alternative to modifications of gravity or exotic particles beyond the standard model. For the effect of primordial black holes on structure formation see also Ref.~\cite{Khlopov:2002yi,Khlopov:2004sc,Khlopov:2008qy}. Moreover, the PBH might also serve as the seeds for the Supermassive Black Holes (SMBH) located in the centers of the galaxies~\cite{Kawasaki:2012kn,Clesse:2015wea}, as well as providing coherent explanations for a host of other problems in the standard cosmological model, such as the too-big-to-fail and the missing-satellite problems~\cite{Garcia-Bellido:2017fdg}.

In the case of micro-clustered PBH, as proposed in Ref.~\cite{Clesse:2015wea}, one would expect that a large fraction of BH encounters will not end up generating bounded systems, but instead just a hyperbolic encounter with emission of GW bremsstrahlung. This is what actually happens if the relative distance or velocity of the two bodies is too high for a BH capture. This case has already been studied in the past in Refs.~\cite{OLeary:2008myb}, and~\cite{Capozziello:2008ra,DeVittori:2012da}, for parabolic and hyperbolic encounters respectively. These events produce enormous {\em bursts} of gravitational waves, which can in principle be bright enough to be detected at cosmological distances \cite{Berry:2013poa,Berry:2013ara}, see also \cite{Kocsis:2006hq} for detection rate estimates from parabolic encounters. The specific waveform of the GW event can be obtained analytically, without the need for computationally-expensive numerical relativity codes, and used to match the coincident event in the three LIGO+Virgo detectors.

It should be noted, however, that the full expression for the power emitted in GW, i.e. the Fourier spectrum of the GW emission, in the case of a hyperbolic encounter given by Eq.~(3.11) of Ref.~\cite{DeVittori:2012da} is incorrect.\footnote{This can easily be
checked, since for eccentricity $e=2$ there is an unphysical pole in the emitted power.} This minor error seems to have escaped the authors as their expression gives the correct limit in the case of parabolic orbits, and in their paper they only considered eccentricities higher than two, due to the choice of the parameters for the astrophysical system considered. We will present the correct expressions, which also have all the correct limits and no singularities, in the Appendix.

Furthermore, in the micro-clustered PBH scenario~\cite{Garcia-Bellido:2017fdg}, we note that for hyperbolic encounters, the characteristic time parameters, as well as the waveform, of the GW emission are very different from those of inspiralling PBH binaries. Furthermore, they provide complementary information which can in principle be used to break degeneracies and estimate the mass, spin and spatial distribution of PBHs as a function of redshift.

The main characteristic of hyperbolic encounters, and the main reason they could be so useful, is that they are one-time events where the main bulk of the energy is released near the periastron. These events also have a uniquely identifiable peak frequency which depends on just the total mass of the system $M$, the relative velocity $v_0$ and the impact parameter $b$. It should be noted that inspiralling and merging PBH has already been studied in the literature, see for example Refs.~\cite{Clesse:2016vqa,Clesse:2016ajp}, where the estimated rate was in the range of tens of events/year/Gpc$^3$ for $M_{\rm PBH} \sim {\cal O}(10-100)\,\Msun$. In Ref.~\cite{Garcia-Bellido:2017qal} is was shown that, within the range of parameters of the micro-clustered PBH scenario~\cite{Clesse:2015wea,Clesse:2016vqa}, the rate of GW burst events in the millisecond range is of the same order of magnitude, if somewhat lower. Nevertheless, for a large fraction of hyperbolic events, the maximum amplitude is well within the noise of the LIGO detectors and without a proper waveform analysis will be very difficult to detect, while in BH spiralling events due to its periodic nature an long duration are much easier to detect.

As described in Ref.~\cite{Garcia-Bellido:2017qal}, hyperbolic encounter events would be detected by future gravitational wave experiments as  bursts with a characteristic frequency at peak strain amplitude. Actually, AdvLIGO has already reported a few events of this type, which were then attributed to noise in the detectors~\cite{Abbott:2016ezn}. However, events from hyperbolic encounters of PBH create shapes similar to the ``tear drop glitch" analyzed in Ref.~\cite{Powell:2016rkl}. 

Therefore, in this paper we continue our analysis of those events, under the assumption that they are actually PBH hyperbolic encounters as their time-frequency profiles could shed light in the  understanding of the AdvLIGO glitches. Finally, if it turns out that the glitches currently observed in AdvLIGO indeed originate from PBH hyperbolic encounters, then this fact could be used to determine the parameters describing the PBHs themselves, i.e. their spatial distribution, velocity and mass.

The layout of our manuscript is as follows:
In Sec.~\ref{sec:relations} we discuss and review the basic relations that determine the geometry and physics of hyperbolic encounters, while in Sec.~\ref{sec:freqpower} we present the corrected expressions for the power spectrum for the emission but also the new analytic expressions for the frequency at peak amplitude. In Sec.~\ref{sec:params} we present the observables that could be extracted from this system while 
we summarize and present our conclusions in Sec.~\ref{sec:conclusions}.

\section{Basic relations and constraints \label{sec:relations}}

\begin{figure}[!t]
\centering
\includegraphics[width = 0.48\textwidth]{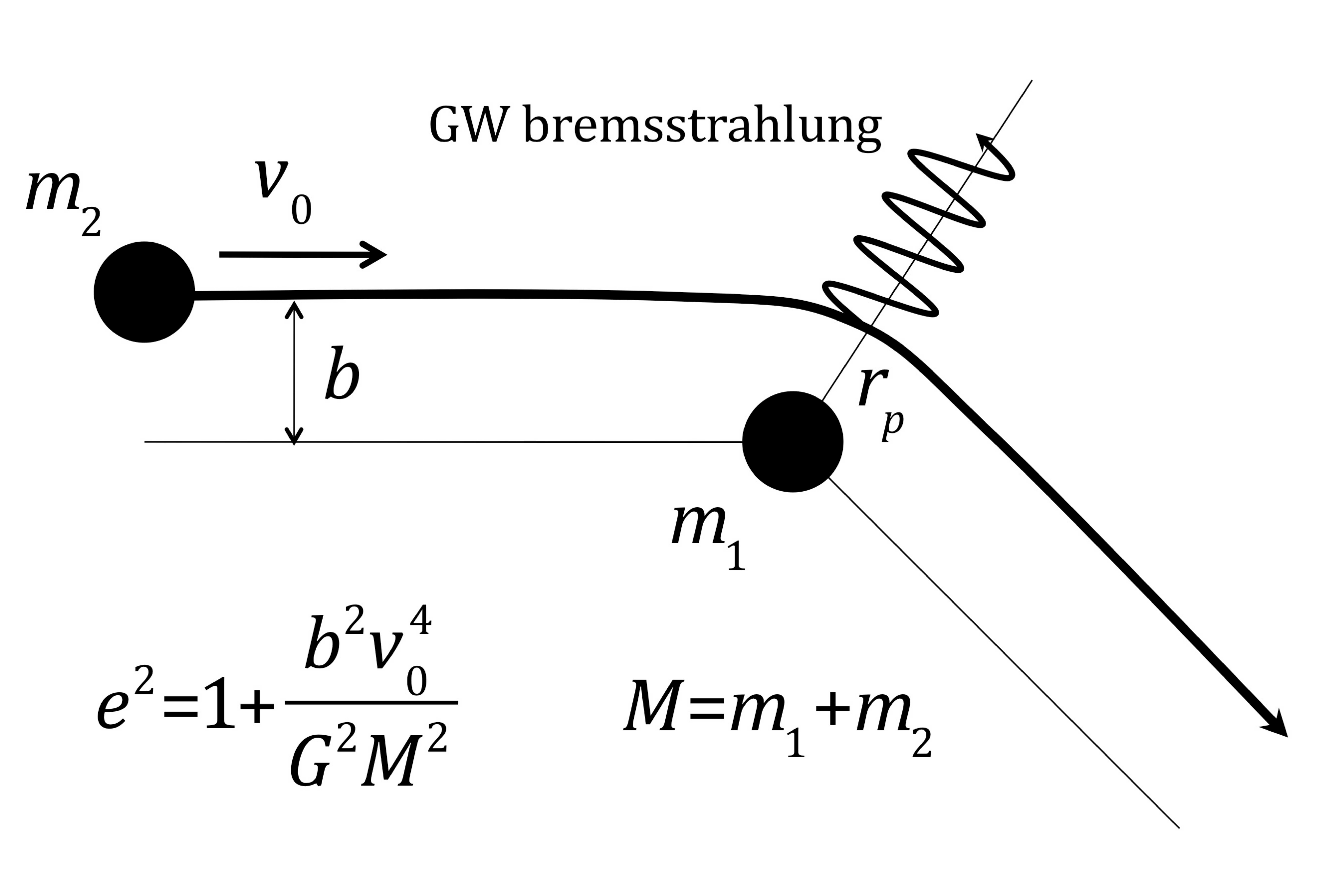}
\caption{The scattering of one BH of mass $m_2$ on another of mass $m_1$ induces the emission of gravitational waves which is maximal at the point of closest approach, $r_p$.}
\vspace*{-1mm}
\label{fig:hyperbolic}
\end{figure}

Consider a hyperbolic encounter between a massive body $m_2$ with asymptotic velocity $v_0$ against a compact mass $m_1$. The total mass is given by $M=m_1+m_2$ and the reduced mass is $\mu = m_1 m_2/M$. Let us assume an impact parameter $b$ as in Fig.~\ref{fig:hyperbolic}. Then, the eccentricity of the hyperbolic orbit is given by
\be\label{eab}
e \equiv \sqrt{1+\frac{b^2}{a^2}} = \sqrt{1+\frac{b^2v_0^4}{G^2M^2}} \ > 1\,.
\ee
The orbital trajectory is characterized in polar coordinates by
\be\label{rphi}
r(\varphi) = \frac{b\,\sin\varphi_0}{\cos(\varphi - \varphi_0) - \cos\varphi_0} =
\frac{a\,(e^2-1)}{1+e\cos(\varphi - \varphi_0)}\,,
\ee
where the relation between the impact parameter $b$ and the semimajor axis $a$ is given by (\ref{eab}), and the angle $\varphi_0$ is given by
\be
\varphi_0 = {\rm arccos}\left(-\frac{1}{e}\right)\,,
\ee
while the distance of maximum proximity is given by
\be\label{rmin}
r_{\rm min} = a\,(e-1) = b\,\sqrt{\frac{e-1}{e+1}} \ > R_s \equiv \frac{2GM}{c^2}\,.
\ee
Conservation of angular momentum implies $b\,v_0 = r_{\rm min}\,v_{\rm max}$. We must impose that $v_{\rm max} < c$ or
\be
\beta \equiv \frac{v_0}{c} < \sqrt{\frac{e-1}{e+1}}\,,
\ee
which substituted into (\ref{eab}) gives
\be
b > R_s\,\frac{(e+1)^{3/2}}{2(e-1)^{1/2}}\,,
\ee
which is a factor $(e+1)/2$ stronger than (\ref{rmin}).

\subsection{Amplitude and power emitted in GW}

\begin{figure}[!t]
\vspace*{2mm}
\centering
\includegraphics[width = 0.49\textwidth]{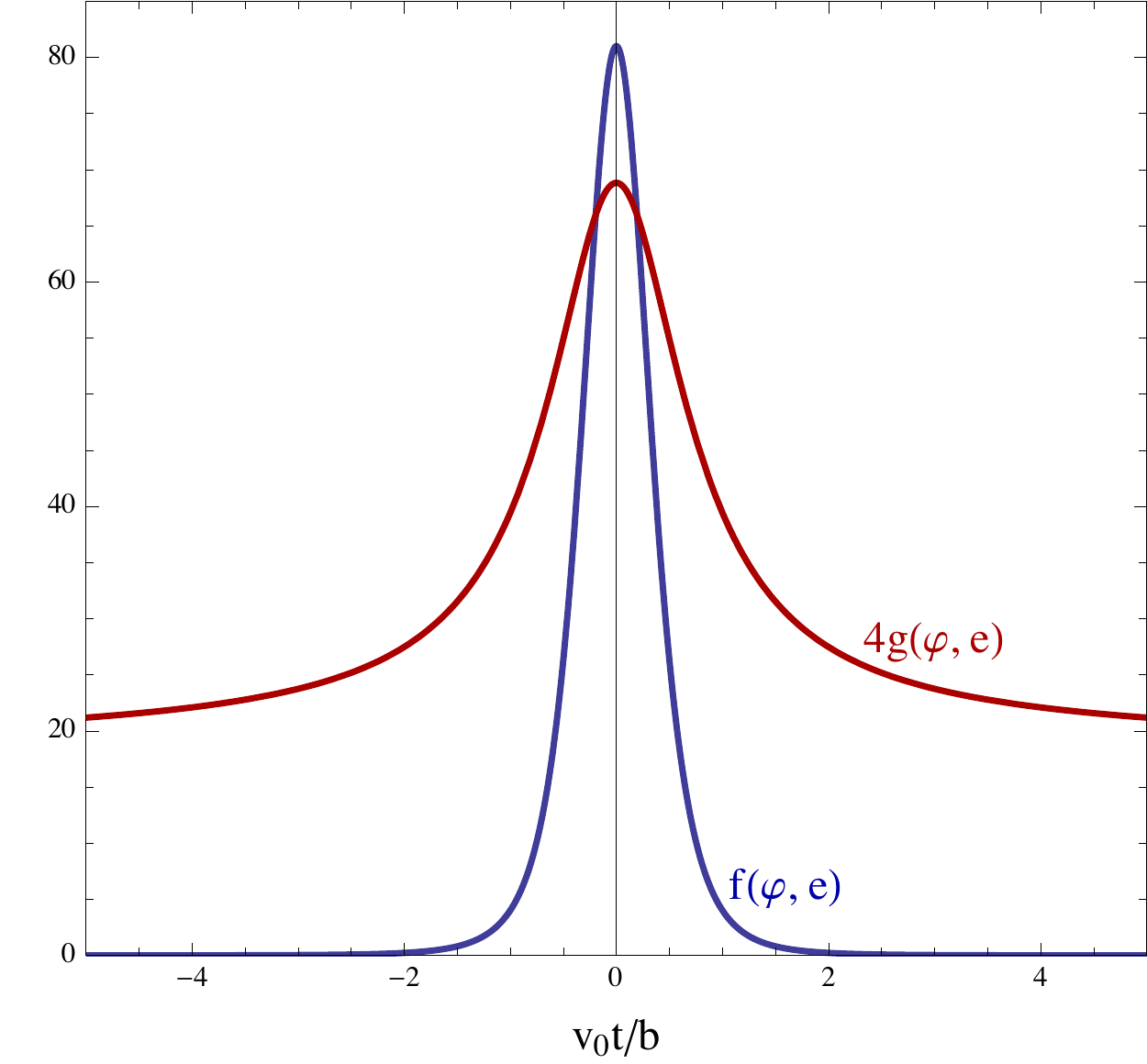}
\caption{The time dependence of the emitted power and strain amplitude of GW in hyperbolic encounters, for the case $e=2$.}
\label{fig:fge}
\end{figure}

The reduced quadrupole moment of the system is given by
\be
Q_{ij} = \mu \,r^2(\varphi) \left(
\begin{array}{l c r}
3\cos^2\varphi - 1 & \ 3\cos\varphi\sin\varphi \ & 0 \\[2mm]
3\cos\varphi\sin\varphi & 3\sin^2\varphi - 1 & 0 \\[2mm]
0 & 0 & -1
\end{array}
\right)\,.
\ee
and the power emitted in GW is then given by
\be\label{eq:power}
P = \frac{dE}{dt} = - \frac{G}{45c^5}\langle \stackrel{\cdots}{Q}_{ij}
\stackrel{\cdots}{Q}{}^{\!ij}\rangle = \frac{32G\mu^2v_0^6}{45c^5\,b^2}\,
f(\varphi,\,e)\,,
\ee
\ba\nonumber
f(\varphi,\,e) &=& \frac{3\left(1+e\,\cos(\varphi-\varphi_0)\right)^4}
{8(e^2-1)^4}\,\Big[24 + 13 \,e^2  \\[2mm] \label{fephi}
&&\hspace*{-5mm} + 48\, e\,\cos(\varphi-\varphi_0)
+ 11\,e^2 \cos2(\varphi-\varphi_0)\Big]
\ea
and the strain amplitude by
\be\label{strain}
h_c = \frac{2G}{R\,c^4}\langle\ddot Q_{ij}\ddot Q^{ij}\rangle^{1/2}_{i,j=1,2}
= \frac{2G\mu \,v_0^2}{R\,c^4}\,g(\varphi,\,e)\,,
\ee
\ba\nonumber
g(\varphi,\,e) &=& \frac{\sqrt2}{e^2-1} \Big[36 + 59 \,e^2 + 10\,e^4
\\[2mm] \label{gephi}
&&\hspace*{-5mm} + (108+47\,e^2)\,e\,\cos(\varphi-\varphi_0) \\[2mm]
&&\hspace*{-5mm} + 59\, e^2\,\cos2(\varphi-\varphi_0)
+ 9\,e^3 \cos3(\varphi-\varphi_0)\Big]^{1/2} \nonumber
\ea
where $f(\varphi,\,e)$ and $g(\varphi,\,e)$ are complicated bell-shaped functions of the angle $\varphi$, symmetric around $\varphi_0$, see Fig.~\ref{fig:fge}. The maximum values occur for $\varphi = \varphi_0$, and only depend on the eccentricity of the orbit,
\ba
f_{\rm max}(e) &=& \frac{9(e+1)^2}{(e-1)^4}\,,\\
g_{\rm max}(e) &=& \frac{2}{e-1}\sqrt{18(e+1)+5e^2}\,.
\ea
The time dependence of these functions can be determined from the relation between angle and time,
\ba\label{time}
t &=& \frac{b}{v_0}\,\int \frac{\sin^2\varphi_0 \,d\varphi}
{(\cos(\varphi - \varphi_0) - \cos\varphi_0)^2} \\[2mm]
&=& \frac{b}{v_0\,e^2}\,\left[\frac{e\,\sin(\varphi - \varphi_0)}{1+e\,\cos(\varphi - \varphi_0)}
-\frac{2}{e+1}\tan\frac{\varphi - \varphi_0}{2}\right] \nonumber \,.
\ea
The functions $f(\varphi,\,e)$ and $g(\varphi,\,e)$ are shown in Fig.~\ref{fig:fge}. The origin of time is chosen to correspond to maximum proximity ($\varphi = \varphi_0$).

\begin{figure*}[!t]
\centering
\includegraphics[width = 0.49\textwidth]{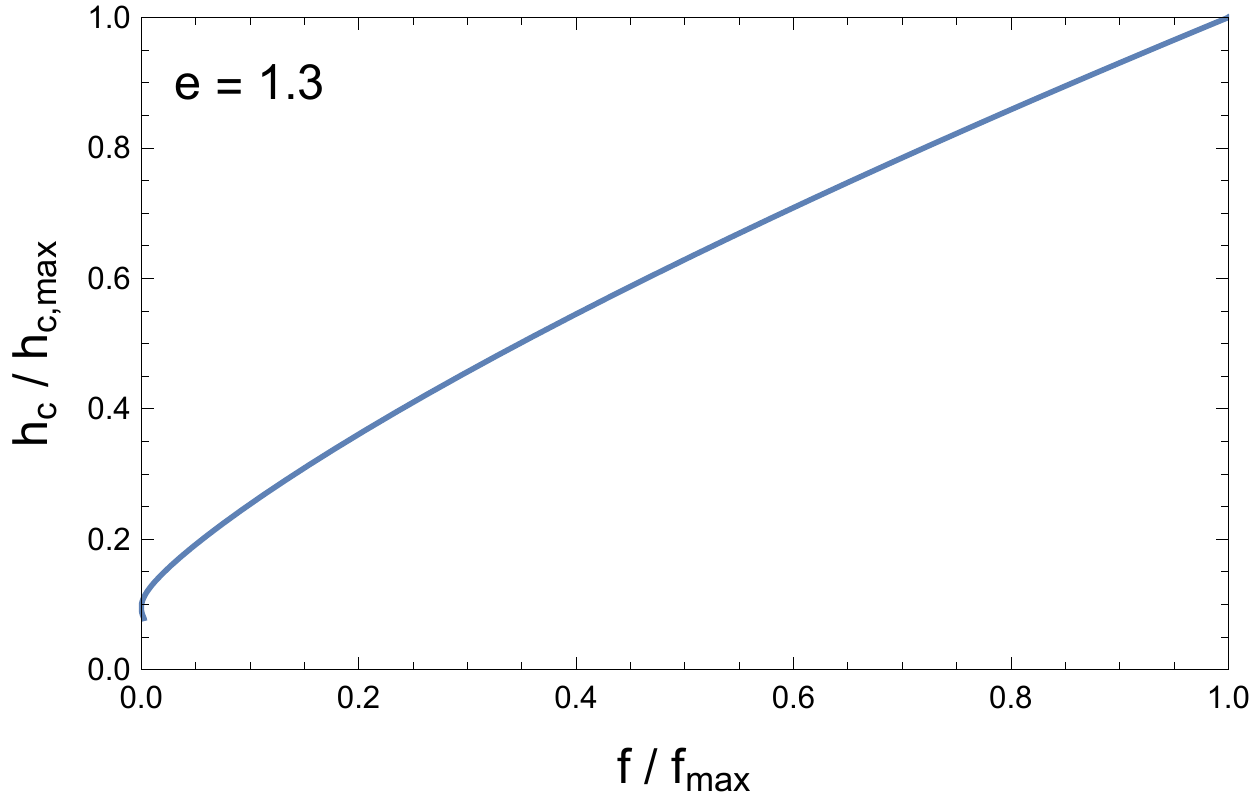}
\includegraphics[width = 0.49\textwidth]{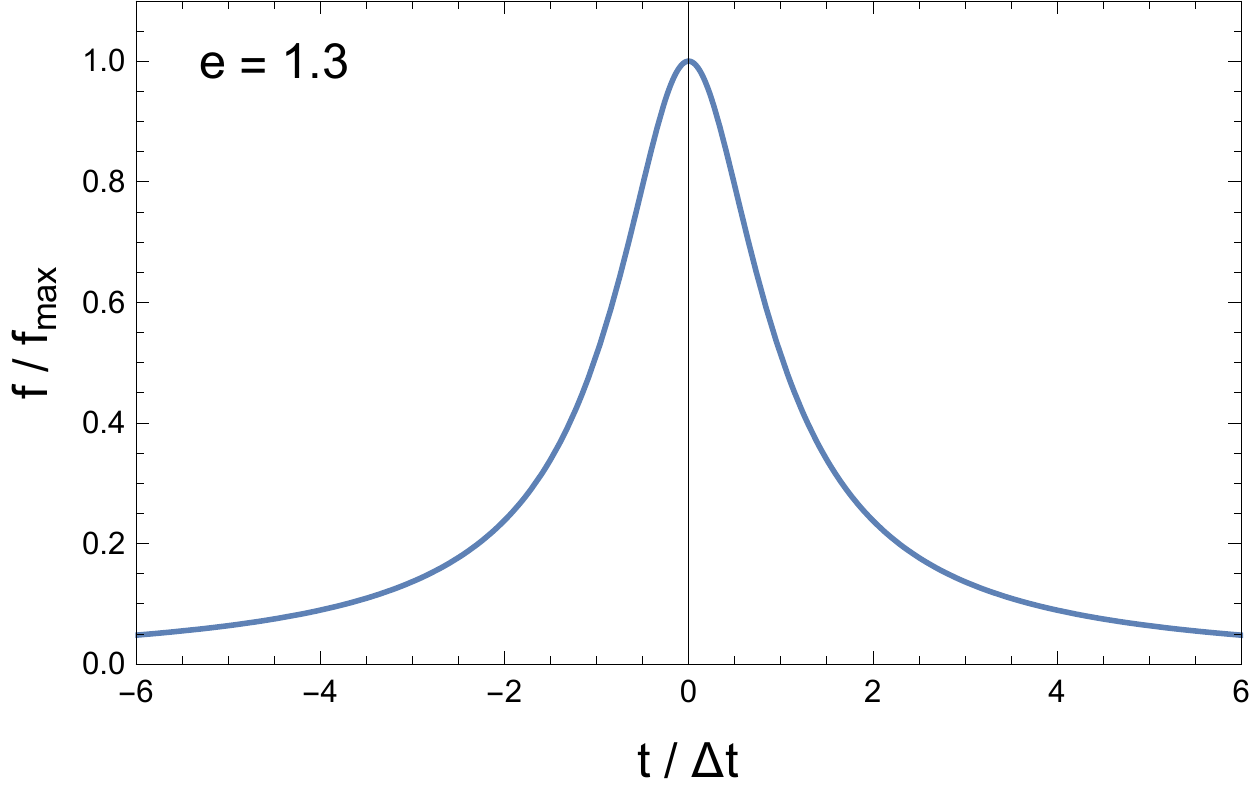}
\caption{The normalized strain $h_c$ as a function of frequency (left) and the frequency in units of the maximum frequency as a function of time (right) for an event with $\beta=0.1, \,M=60\,\Msun, \,b=5\times10^{-5}$ AU and eccentricity $e=1.3$.}
\label{fig:ampfreq}
\end{figure*}

In Fig.~\ref{fig:ampfreq} we show the strain $h_c$ as a function of the frequency $f$ (left) and the frequency as a function of time $t$ (right) for an eccentricity of $e=1.3$. Specifically, we have used $2\pi f = \omega$ and the formulas
\be
\frac{f}{f_{\rm max}} = \left(\frac{1+ e \cos(\phi-\phi_0)}{1+e}\right)^2,
\ee
and
\be
\frac{h_c}{h_{c,{\rm max}}} = \frac{g(\phi,e)}{g_{\rm max}(e)}.
\ee

\subsubsection{LIGO range}

Let us take, for example, the hyperbolic encounter of a black hole of mass $30~\Msun$ moving at $v_0 = 0.01\, c$ towards another black hole of the same mass, with impact parameter $b=10^{-2}$ AU. The eccentricity of the hyperbolic orbit is $e=1.00014$ and the maximum power emitted is given by
\ba\label{Pmax}
P_{\rm max} &=& 2.97\times10^{42}\, {\rm erg/s}
\left(\frac{10^{-2}\, {\rm AU}}{b}\right)^2 \\ \nn
&& \times\left(\frac{\mu}{15\,\Msun}\right)^2
\left(\frac{\beta}{0.01}\right)^6\frac{(e+1)^2}{(e-1)^4}\,.
\ea

In general we have
\be
e^2=1+\left(\frac{b}{10^{-2}\, {\rm AU}}\right)^2
\left(\frac{\beta}{0.01}\right)^4\left(\frac{1\,\Msun}{M}\right)^2
\ee

We can also compute the maximum stress amplitude that such an event would induce on a laser  interferometer on Earth, at a distance $R=1$ Gpc,
\be\label{hmax}
h_c^{\rm max} = 1.63\times10^{-22} \ \frac{\mu}{15\,\Msun}\frac{\rm Gpc}{R}
\left(\frac{\beta}{0.01}\right)^2\frac{g_{\rm max}(e)}{108.5}\,,
\ee
which is perfectly within the sensitivity of LIGO.

The duration of the event can be easily computed from (\ref{time}), as
the time it takes to half the power (\ref{eq:power}) after maximum, and is given by
\be\label{duration}
\Delta t \simeq 50 \,{\rm ms} \ \frac{b}{10^{-2}\, {\rm AU}}\frac{0.01}{\beta}\,\frac{h(e)}{10^{-4}}\,,
\ee
where
\be
h(e)=\frac{2(2^{1/3}-1)(e-1)}{e^2(e+1)^{1/2}(2^{7/6}+(2^{1/3}-1)e-(2^{1/3}+1))^{1/2}}\,.
\ee
The maximum frequency $\omega=2\pi\,f$ is given by
\ba\label{wmax}
\omega_{\rm max}(e) &=& \frac{v_0}{b}\left(\frac{e+1}{e-1}\right) \\ \nn
&=& 20\,{\rm Hz}\ \left(\frac{10^{-2}\, {\rm AU}}{b}\right)
\left(\frac{\beta}{0.01}\right)\left(\frac{e+1}{e-1}\times10^{-4}\right)\,,
\ea
which lies perfectly within the LIGO sensitivity band for close to parabolic encounters.

\begin{figure}[!t]
\vspace*{2mm}
\centering
\includegraphics[width = 0.49\textwidth]{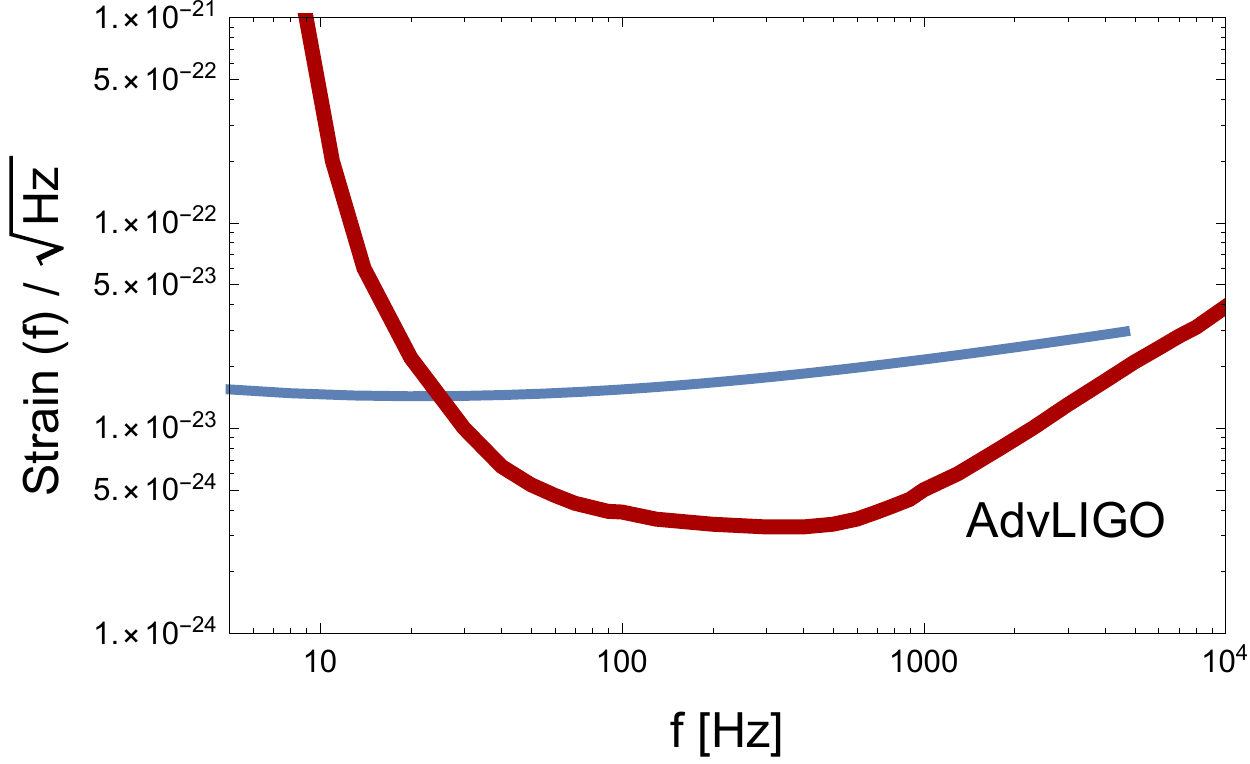}
\caption{The strain $h_c/\sqrt{\rm Hz}$ as a function of frequency in Hz, for $\beta=0.1,\,M=60\,\Msun,\,b=10^{-5}$ AU, which gives
$e=1.014$. Note that the event is fully contained within the AdvLIGO sensitivity, for a maximum frequency $f_{\rm max} = 4.6$ kHz, and a duration of half a millisecond.}
\label{fig:strainfreq}
\end{figure}

We have plotted in Fig.~\ref{fig:strainfreq} the trajectory of a typical event within the AdvLIGO detector, parameterized as the strain $h_c(f)/\sqrt{\rm Hz}$ as a function of frequency, for $\beta=0.1, \,M=60\,\Msun, \,b=10^{-5}$ AU. Note that the event is fully contained within the AdvLIGO sensitivity, for a maximum frequency $f_{\rm max} = 4.6$ kHz, and a duration of half a millisecond.
The event has two stages, the ``chirp" of growing amplitude as the frequency increases to a maximum, and the ``anti-chirp" from this maximum frequency and amplitude to disappearance at low frequencies. Such events should be clearly distinguishable with AdvLIGO+Virgo, searching for bursts in coincidence between the three detectors, within 10 ms and 25 ms respectively.

\subsubsection{LISA range}

Let us consider here an encounter of an IMBH of mass $m_2 = 10^3\,\Msun$ and a SMBH of mass $m_1 = 10^6\,\Msun$. The impact parameter of $b=4$ AU and velocity $v_0 = 0.05\,c$ gives an eccentricity parameter of $e=1.414$ and a maximum power emitted
\ba\label{Pmax2}
P_{\rm max} &=& 4.42\times10^{43} \, {\rm erg/s}
\left(\frac{4\, {\rm AU}}{b}\right)^2 \\ \nn
&& \times\left(\frac{\mu}{10^3\,\Msun}\right)^2
\left(\frac{\beta}{0.05}\right)^6\frac{f_{\rm max}(e)}{1782}\,,
\ea
which is $1.13\times10^{10}$ times larger than the solar luminosity.

In general we have
\be
e^2=1+\left(\frac{b}{1\,{\rm AU}}\right)^2
\left(\frac{\beta}{0.01}\right)^4\left(\frac{10^4\,\Msun}{M}\right)^2
\ee

The maximum stress amplitude that such an event would induce on a laser interferometer on Earth, at a distance $R=1$ Gpc,
\be\label{hmax2}
h_c^{\rm max} = 2.58\times10^{-21} \ \frac{\mu}{10^3\,\Msun}\frac{\rm Gpc}{R}
\left(\frac{\beta}{0.05}\right)^2\frac{g_{\rm max}(e)}{10.34}\,,
\ee
which is perfectly within the sensitivity of LISA.

The duration of the event can be easily computed from Eq.~(\ref{time}), and in this case it is given by
\be\label{duration2}
\Delta t \simeq 11.1 \,{\rm hours} \ \frac{b}{4\, {\rm AU}}\frac{0.05}{\beta}\,h(e)\,,
\ee
and the corresponding maximum frequency is
\be\label{wmax2}
\omega_{\rm max} = 1.4\times10^{-3}\,{\rm Hz}\ \left(\frac{1\, {\rm AU}}{b}\right)^{3/2}
\left(\frac{M}{10^6\,\Msun}\right)^{1/2}
\ee
which lies perfectly within the LISA sensitivity band.

Alternatively, we can consider an encounter between two supergiants of equal masses $m_1=m_2=2\times10^6\,\Msun$, with an impact parameter $b=10$ AU and relative velocity $v_0=0.05\,c$. The eccentricity is low, $e=1.179$, and the stress amplitude is huge
\be\label{hmax3}
h_c^{\rm max} = 2.88\times10^{-18} \ \frac{\mu}{10^6\,\Msun}\frac{\rm Gpc}{R}
\left(\frac{\beta}{0.05}\right)^2\frac{g_{\rm max}(e)}{11.52}\,,
\ee
perfectly detectable by LISA, with a duration of 1.16 days, and a peak power $P_{\rm max} \simeq 6.4\times10^{55}$ erg/s, at $\omega_{\rm max} = 1.41\times10^{-4}$ Hz, right in the middle of LISA sensitivity. Such an event would be clearly distinguishable.

\section{Frequency domain \label{sec:freqpower}}

In this section we present the corrected expressions for the power spectrum for the emission, with respect to those of Ref.~\cite{DeVittori:2012da} but also the new analytic expressions for the frequency at peak amplitude. The main results are presented below, but some useful formulas and more complicated proofs are given in Appendix \ref{sec:app}.

The power spectrum can be obtained from the Fourier transform of the energy emission in the time domain, after also taking into account Parceval's theorem (see Appendix \ref{sec:app}):
\be\label{eq:DE}
\Delta E = \int_{-\infty}^{\infty} P(t)\,dt = \frac{1}{\pi}\int_0^\infty P(\omega)\,d\omega.
\ee

In the case of hyperbolic encounters between two bodies with masses $m_1$, $m_2$ the total energy released via gravitational waves is given by \cite{DeVittori:2012da}:
\be
\Delta E=-\frac{8}{15} \frac{G^{7/2}}{c^5}\frac{M^{1/2} m_1^2 m_2^2}{r_{min}^{7/2}}f(e),\label{eq:enloss1}
\ee
where $r_{min}$ is the minimum distance of the encounter and is related to the impact parameter and the relative velocity at infinity $v_0$ by
\be
b^2 = r_{\rm min}^2 \left(1 + \frac{2GM}{v_0^2 \, r_{\rm min}}\right)\,,
\ee
Then, the factor $f(e)$ and the initial eccentricity of the system are given  by \cite{DeVittori:2012da}
\ba
f(e) &=& \frac{1}{(1+e)^{7/2}} \left[24 \arccos \left(-\frac{1}{e}\right)\left(1+\frac{73}{24}e^2+\frac{37}{96}\right) \right.\nn \\ \label{eq:enlossfactor}
&& + \left.\sqrt{e^2-1}\left(\frac{301}{6}+\frac{673}{12}e^2\right)\right]\,,
\ea
and
\be
e = \sqrt{1+\frac{2 E L^2}{G^2 \mu (m_1 m_2)^2}} \,.
\ee
However, after the close encounter, the system will lose energy and angular momentum due to GW emission, so the eccentricity before and after the encounter will be different. If we assume that the initial energy and angular momentum are $E_{i}=\frac12 M \eta v_0^2$ and $L_{i}=M \eta b v_0$ respectively, where $\eta=\mu/M$, then the final energy and angular momentum after the encounter will be $E_f=E_{i}+\Delta E$ and $L_f=L_{i}+\Delta L$. However, in Ref. \cite{OLeary:2008myb} it was shown that the angular momentum loss $\Delta L$ is proportionally much smaller than the energy loss, i.e. $\Delta L/L_{i}\ll \Delta E/E_{i}$, so in what follows we will ignored it. Therefore, the initial and final eccentricities, $e_i$ and $e_f$, will be given by
\be
e_i^2 - 1 = \frac{2 E_{i} L_{i}^2}{G^2 \mu (m_1 m_2)^2}
= \frac{v_0^4 \,b^2}{G^2 M^2}\,,
\ee
and
\be
e_f^2 - e_i^2 = \frac{2 b^2 v_0^2 \Delta E}{G^2 M^2 \eta}\,.
\ee

The expression of the energy power spectrum of the GW emission, as calculated in Ref.~\cite{DeVittori:2012da}, is incorrect since we have found that there is an unphysical pole for $e=2$ due to the $ive/2$ terms in their expressions. Also, the integral of the power over all frequencies does not give the exact analytical result~\cite{DeVittori:2012da}, so in what follows we present the correct expressions. From Eq.~(\ref{eq:DE}) we can see that in Fourier space the power is given by
\ba
P(\omega)&=&\frac{G}{45 c^5}~\sum_{i,j} |\widehat{\dddot{Q}_{ij}}|^2\nn\\
&=&\frac{G}{45 c^5}~\omega^6 \sum_{i,j} |\widehat{Q_{ij}}|^2,
\ea
where $\widehat{Q_{ij}}$ is the Fourier transform of the quadrupole momentum tensor $Q_{ij}$ which is given in terms of the variable $\xi$ by
\begin{widetext}
\be
Q_{ij}=\frac12 a^2 \mu\left(
         \begin{array}{ccc}
           (3-e ^2) \cosh 2 \xi -8 e  \cosh \xi  & 3 \sqrt{e ^2-1} (2 e  \sinh \xi -\sinh 2 \xi ) & 0 \\
           3 \sqrt{e ^2-1} (2 e  \sinh \xi -\sinh 2 \xi) & (2 e ^2-3) \cosh 2 \xi +4 e  \cosh \xi  & 0 \\
           0 & 0 & 4 e\cosh \xi-e^2  \cosh 2 \xi  \\
         \end{array}
       \right),
\ee
\end{widetext}
where we have dropped some constant terms as they do not affect the quadrupole tensor, as we can always absorb them by making a translation, and we have that the variable $\xi$ is related implicitly to the time and radial coordinates, for $\nu_0 = \sqrt{a^3/G M}$, by
\ba
t(\xi) &=& \nu_0 (e \sinh\,\xi-\xi)\,, \\[1mm]
r(\xi) &=& a (e \cosh\,\xi-1)\,.
\ea

\begin{figure*}[!t]
\centering
\includegraphics[width = 0.49\textwidth]{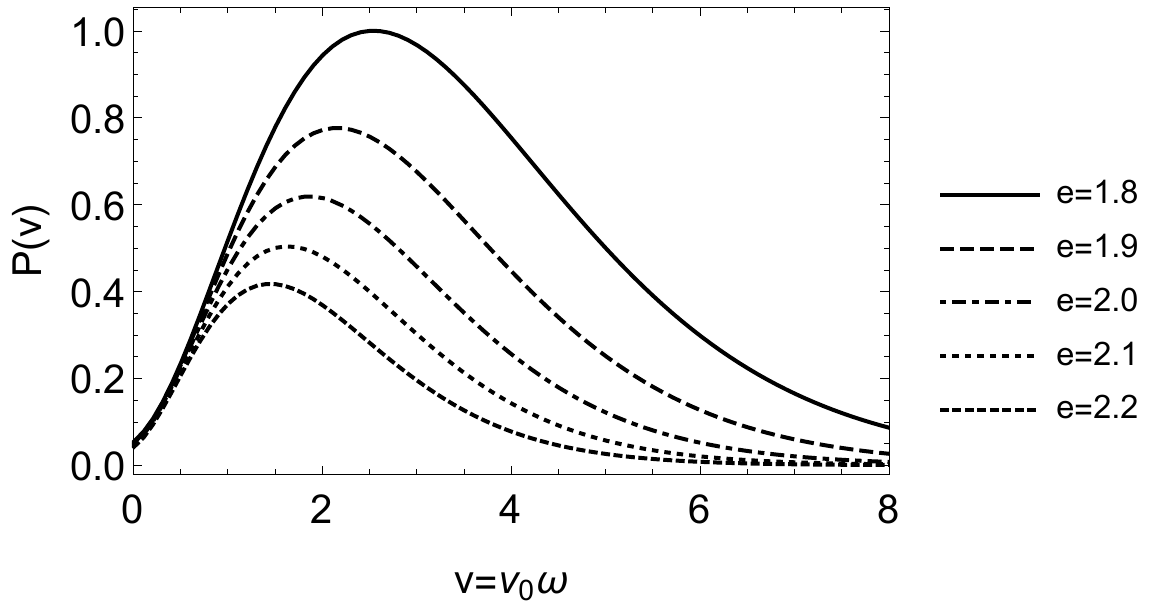}
\includegraphics[width = 0.49\textwidth]{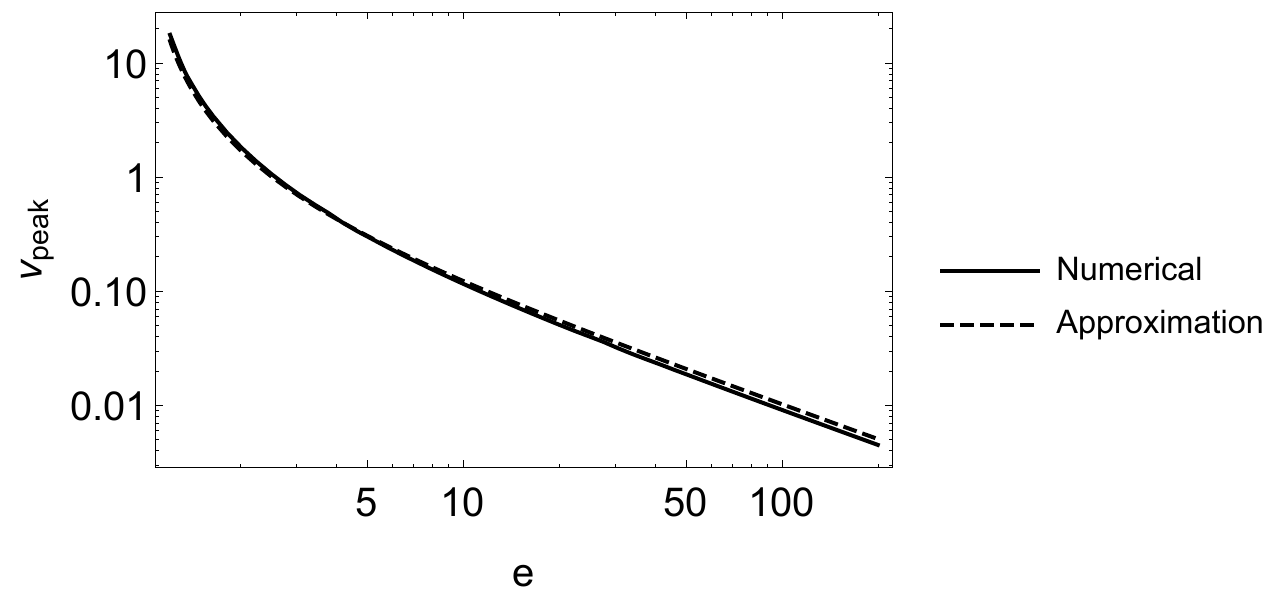}
\caption{Left: The power spectrum as a function of the dimensionless variable $\nu=\nu_0 \omega$ for various values of the eccentricity e. We have normalized the power with respect to that of $e=1.8$. As can be seen, the power does ont have a pole at $e=2$, but also is non-zero for $v=0$. Right: Comparison of the approximate vs numerical value of the frequency at peak power. Clearly, the agreement is excellent.}
\label{fig:plots1}
\end{figure*}

The Fourier transform of $Q_{ij}$, i.e. $\widehat{Q_{ij}}$, is obtained by using the corresponding expressions for $\sinh\,\xi$ etc given in Appendix \ref{sec:app}.
Setting $\nu=\nu_0\, \omega$ and
\be
\widehat{C_{ij}}=\left(\frac{a^2 \mu}2  \frac{\pi}{\omega}\right)^{-1}\widehat{Q_{ij}}\,.
\ee
we can then write the power in term of the dimensionless frequency $\nu$ as:
\ba
P(\omega)&=&\frac{G^3 \mu^2M^2}{a^2 c^5}~\left(\frac{\pi^2}{180}~\nu^4 \sum_{i,j} |\widehat{C_{ij}}|^2\right)\nn \\
&=& \frac{G^3 \mu^2M^2}{a^2 c^5}~\frac{16\pi^2}{180}~\nu^4 F_e(\nu),
\ea
where we have defined
\ba\nn
F_e(\nu)&=& \left|\frac{3(e^2-1)}{e}\,H_{i \nu}^{(1)}{}'(i\nu e)+\frac{e^2-3}{e^2}\,
\frac{i}{\nu}\,H_{i\nu}^{(1)}(i\nu e)\right|^2  \\
&+& \left|\frac{3(e^2-1)}{e}\,H_{i\nu}^{(1)}{}'(i\nu e)+\frac{2e^2-3}{e^2}\,
\frac{i}{\nu}\,H_{i\nu}^{(1)}(i\nu e)\right|^2 \nn \\
&+& \left|\frac{i}{\nu}\,H_{i \nu}^{(1)}(i\nu e)\right|^2 \label{Fe} +
\frac{18\,(e^2-1)}{e^2}\times\nn \\
&\times& \left|\frac{(e^2-1)}{e}\,i\,H_{i\nu}^{(1)}(i\nu e)+
\frac{1}{\nu}\,H_{i \nu}^{(1)}{}'(i\nu e)\right|^2\,,
\ea
Then, the total energy is
\ba
\Delta E&=& \int_{-\infty}^{+\infty} P(t) dt  =
\int_{0}^{+\infty} \frac{P(\omega)}{\pi} d\omega \nn\\
&=& \left(\frac{G^{7/2} \mu^2 M^{5/2}}{c^5 a^{7/2}}\right)  \frac{16\,\pi}{180}\int_{0}^{+\infty}\!\! \nu^4 F_e(\nu) d \nu\,. \label{eq:deltaE1}
\ea
This expression can then be compared with the corresponding one in the time domain given by Eq.~(\ref{eq:enloss1}) with the eccentricity dependence given by Eq.~(\ref{eq:enlossfactor}). We have tested by numerically integrating Eq.~(\ref{eq:deltaE1}) that both Eq.~(\ref{eq:enloss1}) and (\ref{eq:deltaE1}) agree exactly for all values of the parameters. Also, our expressions agree with in the parabolic limit ($e=1$) with those of Refs.\cite{Turner:1977tm,Berry:2010gt}.

The power can be approximated at high frequencies and small eccentricities by
\ba
\nu^4 F_e(\nu) &\simeq& \frac{12\,F_y(\nu)}{\pi \,y \,(y^2+1)^2}
\,e^{-2\nu z(y)} \,, \\ \nn
F_y(\nu) &=& \nu\left(1-y^2-3\nu y^3+4y^4+9\nu y^5+6\nu^2y^6\right)\,,\\
z(y) &=&y - {\rm arctan}\,y \,,\hspace{5mm}
y \equiv \sqrt{e^2-1}\,,
\ea
which has a maximum at
\be
\nu_{\rm max}(e) = \sqrt\frac{e+1}{(e-1)^3}\,, \hspace{3mm} \label{eq:vmax}
\omega_{\rm max}(e) = \frac{v_0}{b}\left(\frac{e+1}{e-1}\right)\,.
\ee
It is easy to check that the maximum power in the time and frequency domains coincide numerically in the whole range of eccentricities $e>1$.

Using (9.3.15 - 9.3.20) of Abramowitz \& Stegun \cite{Abramowitz:1974} we can write
\ba
H_{i \nu}^{(1)}(i\nu e) &\simeq& \frac{2(e^2-1)^{-1/4}}{i\,\sqrt{2\pi\nu}}\,e^{-\nu z(e)}\\
H_{i \nu}^{(1)}{}'(i\nu e) &\simeq& \frac{2(e^2-1)^{1/4}}{e\,\sqrt{2\pi\nu}}\,e^{-\nu z(e)}
\ea
Then, the integral over frequencies is
\ba
\int_0^\infty \!\!\nu^4 F_e(\nu) d\nu &\simeq& \frac{y(1275+673y^2)}{2\pi\,y^7}\\
 &+& \frac{3(425+366y^2+37y^4){\rm arctan}\,y}{2\pi\,y^7} \,. \nn
\ea

By taking the limit $\nu \to 0$,
\ba
\lim_{\nu\to0}H_{i \nu}^{(1)}(i\nu e) &=& \frac{2i}{\pi}\ln(\nu e)\,,\\
\lim_{\nu\to0}H_{i \nu}^{(1)}{}'(i\nu e) &=& \frac{2}{\pi\nu e}\,,
\ea
we can find
\be
\lim_{\nu\to0} \nu^4\,F_e(\nu) = \frac{72(e^2-1)}{\pi^2e^4}\,,
\ee
which is finite and different from zero, except for $e=1$ and $e\to\infty$.
Then, the power for zero frequency becomes:
\be
P(\omega=0)=\frac{G^3 \mu^2M^2}{a^2 c^5}~\frac{32 \left(e^2-1\right)}{5 e^4} \label{eq:memory}.
\ee
This result implies that even in the limit of zero frequency there is energy emitted by the system, which is in fact in agreement with Ref.~\cite{DeVittori:2014psa}, where it was shown that the cross-polarization state of GWs emitted by nonspinning compact binaries in hyperbolic orbits exhibits a memory effect for GWs, i.e. there is a non zero difference in the amplitude and hence energy emitted between the two states at $t=\pm \infty$. This fact is another reason of disagreement between our new expressions and those of Ref.~\cite{DeVittori:2012da} which for zero frequency predict no emitted power. When the eccentricity goes to either one or infinity, then the emitted power for zero frequency is again zero as expected.

In the left panel of Fig.~\ref{fig:plots1} we show the power spectrum as a function of the dimensionless variable $\nu=\nu_0 \omega$ for various values of the eccentricity e. We have normalized the power with respect to tat of $e=1.8$. As can be seen, the power does ont have a pole at $e=2$, but also is non-zero for $v=0$. On the right panel we show a comparison of the approximate vs numerical value of the frequency at peak power, with the approximate given by Eq.~(\ref{eq:vmax}). Clearly, the agreement between the two cases is excellent.

Finally, we can also define the characteristic strain $h_c$ of the emission via the relation
\be
\frac{dE}{dt}\sim \frac{c^3 d_L^2}{4G}|\dot{h}|^2 \sim \frac{c^5}{4G}\left(\frac{\omega d_L h}{c}\right)^2,
\ee
which implies
\be
h_c(z,\omega)\simeq \frac1{d_L(z)}\sqrt{\frac{4G}{c^3} P(\omega)}.
\ee

\section{Parameters and observables\label{sec:params}}

For GW-bursts observations, there are five independent parameters $(M, \mu, v_0, b, R)$ and a derived one, the eccentricity (\ref{eab}), while there are at least six observables $(\Delta t, \omega_{\rm max}, h_{\rm max}, P_{\rm max}, {\rm shape}\,h_c(t), {\rm shape}\,P(\omega))$. There are therefore more observables than parameters. In particular, there is a relation between observables that allows one to obtain directly the values of some parameters and constraints among them. For instance,
\be
\Delta t\ \omega_{\rm max}(e) = \frac{e+1}{e-1}\,h(e)\,,\label{cons1}
\ee
which allows one to deduce the eccentricity of the event from its duration and its maximum emitted frequency, both well defined observables, and then
\be\label{cons2}
R = c\,\Delta t \left(\frac{45}{8}\frac{G\,P_{\rm max}}{c^5\,f_{\rm max}(e)}
\frac{g^2_{\rm max}(e)}{h^2_{c, {\rm max}}}\right)^{1/2}\,,
\ee
gives the distance to the event. The other observables can be used to determine the rest of the parameters, $(M, \mu, v_0, b)$. For instance
\be\label{cons3}
\frac{b^2}{v_0^2} = \frac{G^2M^2}{v_0^6}(e^2-1)\,,
\ee
and
\be\label{cons4}
\mu v_0^2 = R\,\frac{h_c^{\rm max}}{2G}\frac{c^4}{g_{\rm max}(e)}\,,
\ee
together with
\be\label{cons5}
\frac{GM}{v_0^3} = \frac{b}{v_0\sqrt{e^2-1}}=\frac{\Delta t}{h(e)\sqrt{e^2-1}}\,,
\ee
will allow one to determine the ratio of the two masses $q=m_1/m_2\geq1$, from $M/\mu = (1+q)^2/q$.

Moreover, the rate of events per unit volume can be computed by first considering the individual collision rate
\be
\tau_{\textrm{ind}}=n_{\textrm{PBH}}~v_{\textrm{PBH}}~\sigma,
\ee
where
\ba
n_{\textrm{PBH}}&=&\delta_{\textrm{PBH}}^{local}~\rho_{\textrm{DM}}/M_{\textrm{PBH}}\nn \\
&=&\delta_{\textrm{PBH}}^{\textrm{local}}~\rho_\textrm{c}~\Omega_{\textrm{DM}}/M_{\textrm{PBH}}
\ea
is the number density of PBHs, $\delta_{\textrm{PBH}}^{\textrm{local}}$ is the local density contrast, $\rho_\textrm{c}$ is the critical density and $\Omega_{\textrm{DM}}\simeq0.25$ the dark matter density, while $v_{\textrm{PBH}}$ is the relative velocity of the PBHs and $\sigma=\pi b^2$ the cross-section for an impact parameter $b$. We then find that
\ba
\tau_{\textrm{ind}}&=&1.57\times10^{-23} \left(\frac{\delta_{\rm PBH}^{\rm local}}{10^6}\right) \left(\frac{v_{\textrm{PBH}}}{200\,\textrm{km/s}}\right)\nn\\&&\times\left(\frac{b}{10^{-5}\textrm{AU}}\right)^2\left(\frac{M_{\textrm{PBH}}}{30\,M_{\odot}}\right)^{-1}\textrm{yr}^{-1}
\ea
and the total rate per comoving volume in units of $\textrm{Gpc}^3$ is obtained by multiplying the total number of events in the volume $N=n\,V$ with the individual rate $\tau_{\textrm{ind}}$, or $\Gamma_{\textrm{total}}= n_{\textrm{PBH}}\,\tau_{\textrm{ind}}\,V$. Therefore,
\ba
\Gamma_{\rm total}/V&=& n_{\textrm{PBH}}^2~v_{\textrm{PBH}}~\sigma \nn \\
&=& 16.3 \left(\frac{\delta_{\rm PBH}^{\rm local}}{10^6}\right)^2 \left(\frac{v_{\textrm{PBH}}}{200\,\textrm{km/s}}\right)\left(\frac{b}{10^{-5}\textrm{AU}}\right)^2
\nn\\&&\times \left(\frac{M_{\textrm{PBH}}}{30\,M_{\odot}}\right)^{-2} \textrm{yr}^{-1}\textrm{Gpc}^{-3}.
\ea

This can also be re-written as
\ba
\Gamma_{\rm total}/V &=& \frac{9}{64\pi}\frac{H_0^4}{v_0^3}
\Big(\delta_{\rm PBH}^{\rm local}\,\Omega_{\rm DM}\Big)^2(e^2-1) \nn \\
&=& 25.4\ {\rm yr}^{-1}{\rm Gpc}^{-3}\left(\frac{\delta_{\rm PBH}^{\rm local}}{10^8}\right)^2
\frac{e^2-1}{\beta^3},
\ea
for $h=0.7$ and $\Omega_{\rm DM}=0.25$, which can be significantly large for $\beta\ll1$. Finally, there is also a simple relation for the total power
\be
P_{\rm max} = \frac{32}{45}\frac{q^2\,\beta^{10}}{(1+q)^4}
\frac{9\,(e+1)}{(e-1)^5}\,\frac{c^5}{G}\,,
\ee
in units of $c^5/G = M_P/t_P = 3.6295\times10^{59}$ erg/s = $9.3064\times10^{25}\ {\cal L}_\odot$, which can be very large for close encounters (near-parabolic) and large velocities.


\begin{figure*}[!t]
\centering
\includegraphics[width = 0.49\textwidth]{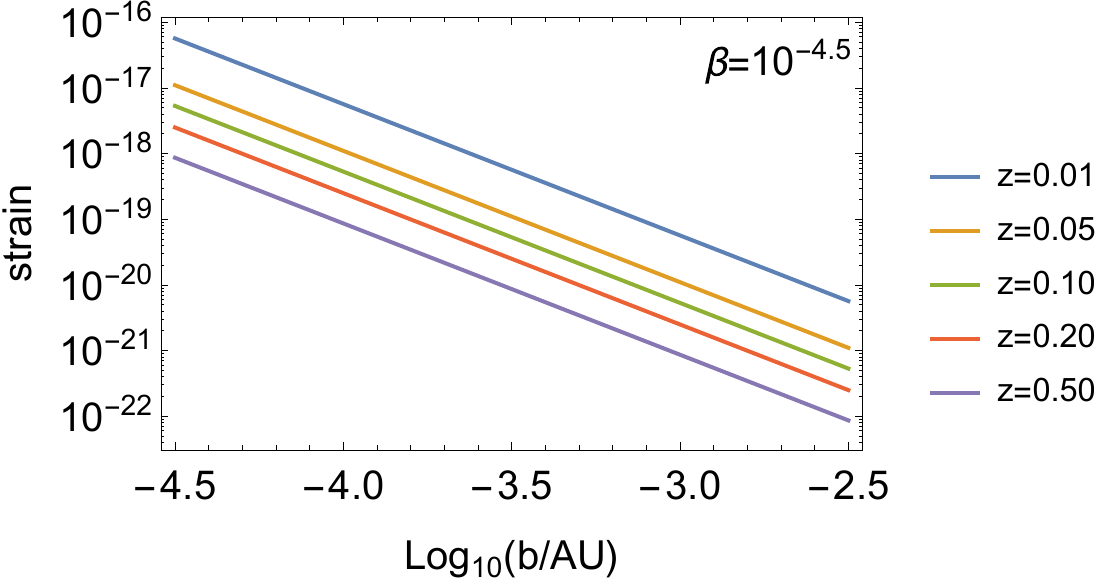}
\includegraphics[width = 0.49\textwidth]{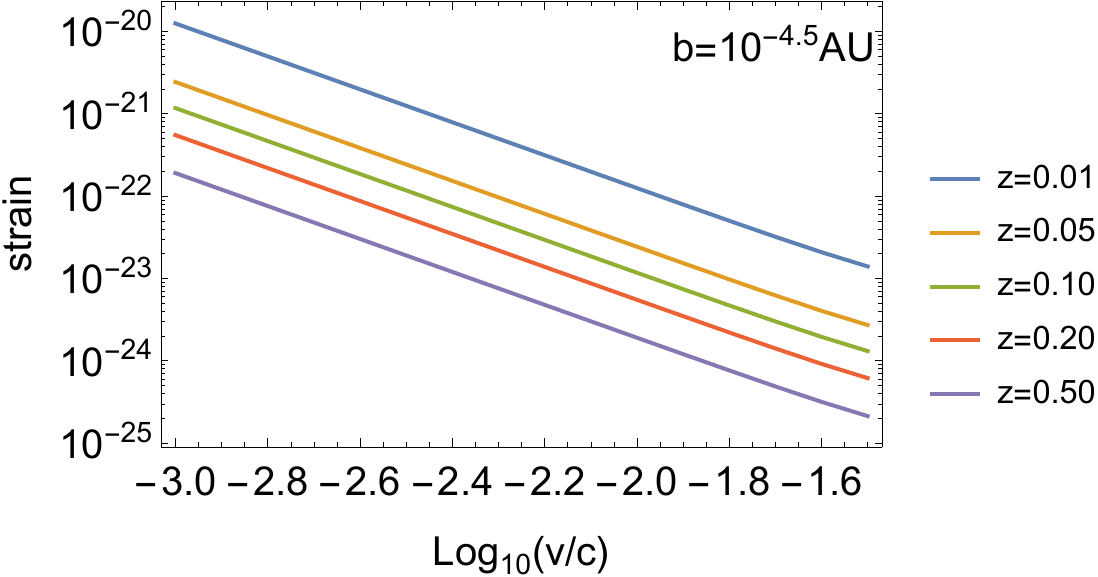}
\caption{The strain at different redshifts for various velocities and impact parameters for a PBH encounter with parameters $M=3 M_{\bigodot}$.}
\label{fig:plots4}
\end{figure*}

\begin{figure}[!t]
\centering
\includegraphics[width = 0.49\textwidth]{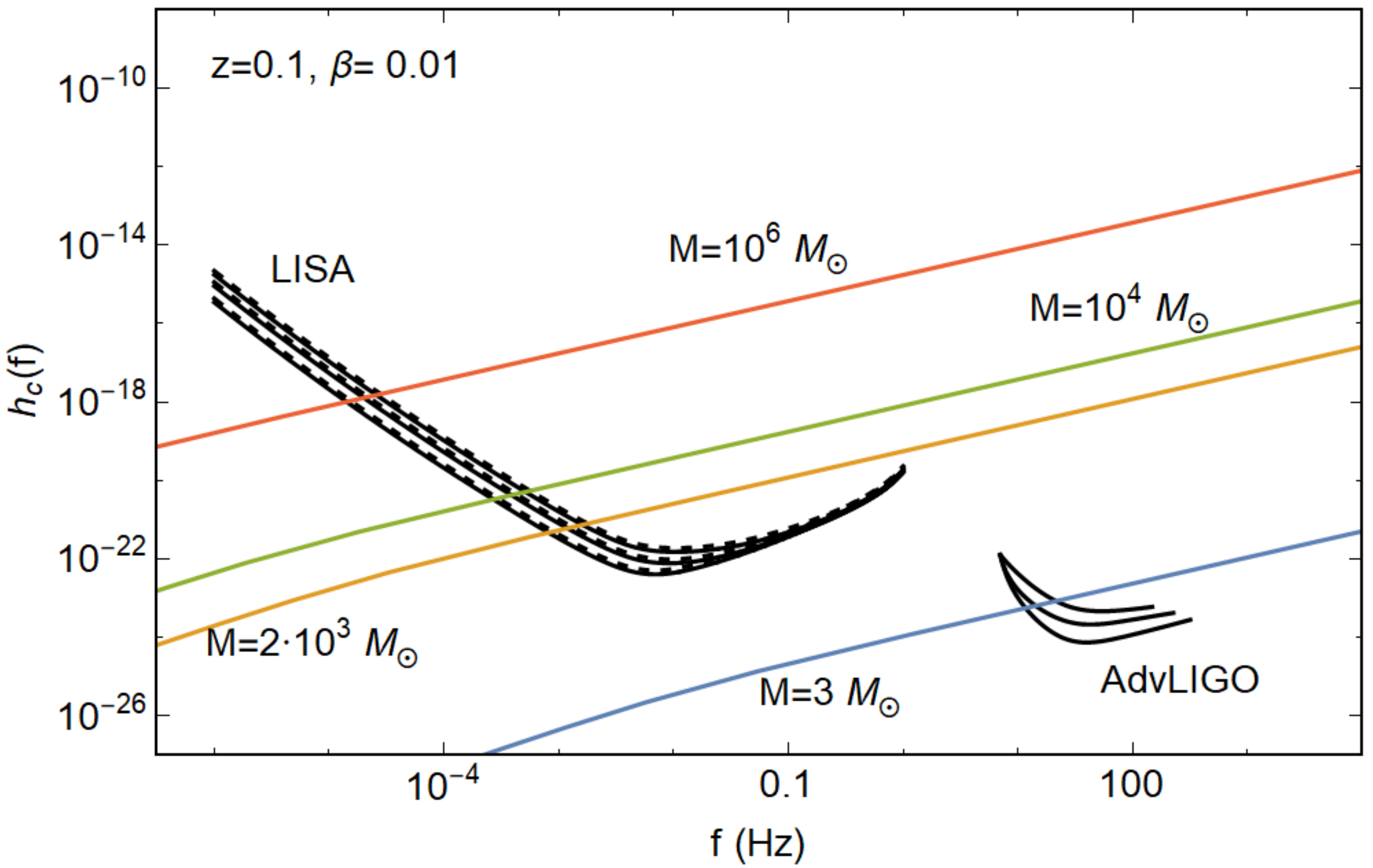}
\caption{The AdvLIGO and LISA sensitivity curves, together with the locus of peak frequencies, as a function of impact parameter, $\log_{10}(b/AU)\in[-6,3]$, for different PBH total masses, $3 - 10^6\,\Msun$, and a redshift to the source of $z=0.1$. The low end of these values is compatible with $\bar\mu = 3\,\Msun$ and $\sigma=0.5$ according to the analysis of Ref.\cite{Clesse:2017bsw}.}
\label{fig:contours1}
\end{figure}

\begin{figure}[!t]
\centering
\includegraphics[width = 0.49\textwidth]{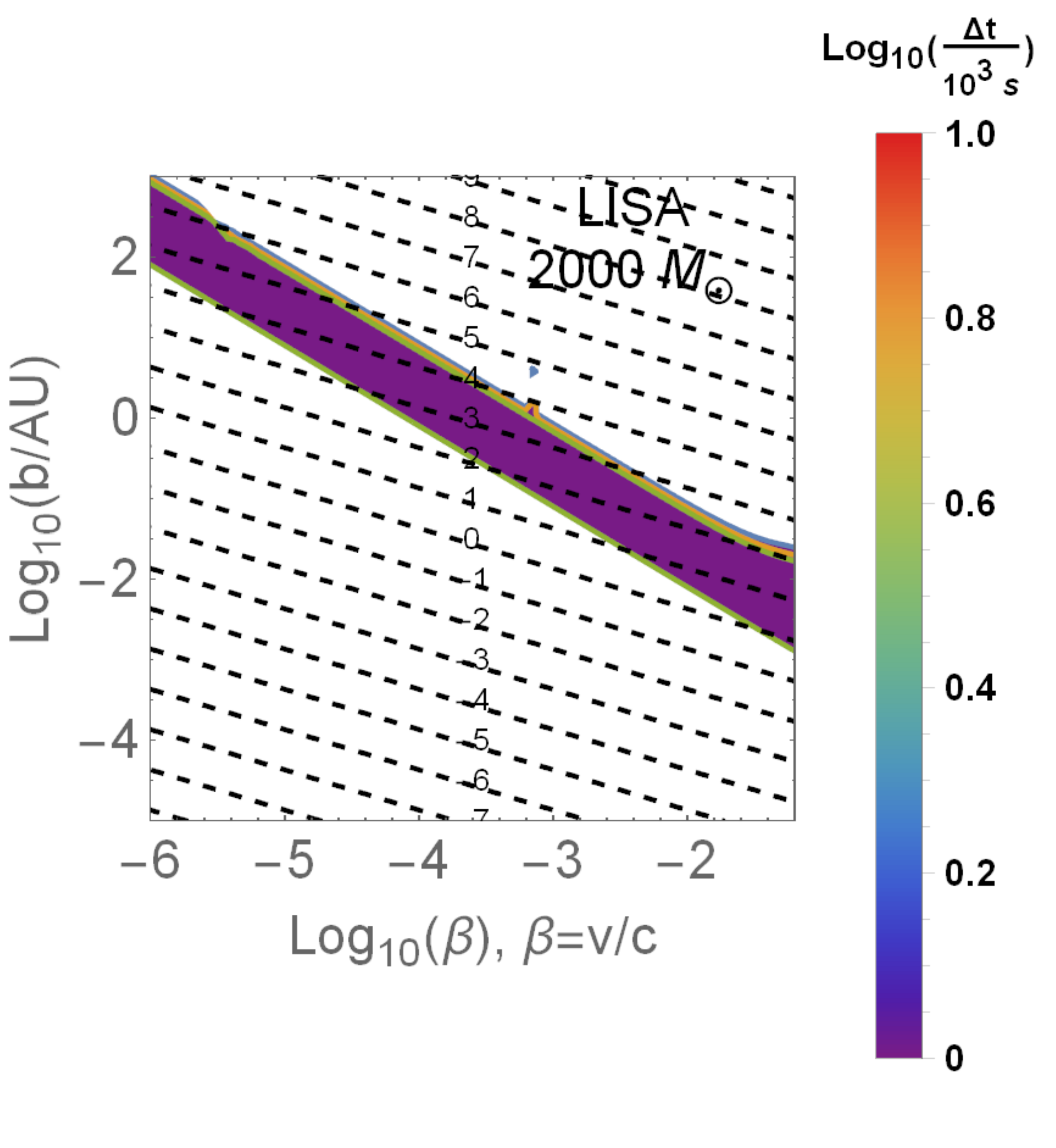}
\caption{The parameter space accessible to LISA sensitivities, in terms of the relative velocity $\beta=v_0/c$ and the impact parameter $b/\textrm{AU}$.
We assume the BH pair to be at a redshift of $z=0.1$ and total mass $M=2000\Msun$. The coloring scheme corresponds to the characteristic timescale of the event, while the dashed lines show the expected event rate in units of events/yr/Gpc$^3$, i.e. $\log_{10}(\Gamma/V)$. The band width is related to the sensitivity of the detectors, see Fig.~\ref{fig:contours1}.}
\label{fig:contours2}
\end{figure}

As mentioned in Sec.~\ref{sec:relations} we also impose constraints on the impact parameter $b$ so that the BHs remain unbound, and the impact parameter is larger than the Schwarzschild radius $R_s = 2GM/c^2$ and we also consider relative speeds $v_0<0.1\,c$, in order to remain in the non-relativistic regime. 

For comparison, in Fig.~\ref{fig:plots4} we show the strain at different redshifts for various velocities and impact parameters for a PBH encounter with parameters $M=3 M_{\bigodot}$, while in Fig.~\ref{fig:contours1} we show the AdvLIGO and LISA sensitivity curves, together with the locus of peak frequencies, as a function of impact parameter, $\log_{10}(b/AU)\in[-6,3]$, for different PBH total masses, $3 - 10^6\,\Msun$, and a redshift to the source of $z=0.1$. Finally, in Fig.~\ref{fig:contours2} we also show the parameter space accessible to LISA sensitivities, in terms of the relative velocity $\beta=v_0/c$ and the impact parameter $b/\textrm{AU}$.
As an example we assume the BH pair to be at a redshift of $z=0.1$ and total mass $M=2000\Msun$. The coloring scheme corresponds to the characteristic timescale of the event, while the dashed lines show the expected event rate in units of events/yr/Gpc$^3$, i.e. $\log_{10}(\Gamma/V)$. The band width is related to the sensitivity of the detectors, see Fig.~\ref{fig:contours1}.

\section{Conclusions \label{sec:conclusions}}

We found that hyperbolic encounters of PBHs with relative velocities of the order of $\sim 0.1c$, at distances of each other $\sim 10^{-4}$ AU and at redshifts in the range $z\in[0,0.5]$ would produce one-time GW bursts with frequency and strains well within the sensitivity of the LISA experiment. These events would have unique signatures, very different from the usual in-spiralling stellar BHs, and would provide strong evidence in favor of the PBH paradigm. In particular, the trajectory of the event within the sensitivity band is almost flat in strain/Sqrt(Hz) as a function of frequency, see Fig.~\ref{fig:strainfreq}, and it has a well defined maximum frequency.

Therefore, it will be possible to start seeing events crossing the sensitivity band and then disappearing into their noise, to later appear again, as the BH moves away from the close encounter. It is a very different waveform from BHB inspirals, one which can in principle be easily detectable and distinguishable as a burst.

Furthermore, in this analysis we present corrected formulas, for the power spectrum in terms of the frequency, of hyperbolic encounters found in the literature and explicitly show new exact and approximate expressions for the peak frequency of the emission. We have tested that the expressions for the power spectrum, when integrated over all frequencies, give the well known result for the energy loss in a hyperbolic encounter.
\\~\\~
\textbf{Numerical Analysis Files}: The numerical codes (Mathematica and Python) used by the authors in the analysis of the paper can be found \href{http://members.ift.uam-csic.es/savvas.nesseris/gws.html}{here}.

\section*{Acknowledgements}
The authors acknowledge support from the Research Project FPA2015-68048-03-3P [MINECO-FEDER] and the Centro de Excelencia Severo Ochoa Program SEV-2016-0597. JGB thanks the Theory Department at CERN for their hospitality during a Sabbatical year at CERN. He also acknowledges support from the Salvador de Madariaga Program Ref. PRX17/00056. S.N. acknowledges support from the Ram\'{o}n y Cajal program through Grant No. RYC-2014-15843.

\appendix
\section{Useful formulae \label{sec:app}}
In our analysis we always follow the notation of Landau and Lifshitz \cite{landau1975classical} for the Fourier transform:
\ba
f(t)&=&\frac{1}{2\pi}\int_{-\infty}^{+\infty} \widehat{f}(\omega) e^{-i\omega t} d \omega \\
\widehat{f}(\omega)&=&\int_{-\infty}^{+\infty} f(t) e^{i\omega t} d \omega \\
\delta(\omega-\omega')&=&\frac{1}{2\pi}\int_{-\infty}^{+\infty} f(t) e^{-i(\omega-\omega') t} d t
\ea

Then, Parceval's theorem can be proven easily as follows:
\ba
\int_{-\infty}^{+\infty} |x(t)|^2 dt&=&\int_{-\infty}^{+\infty}\frac{1}{2\pi} \frac{1}{2\pi}\nn\\
&\cdot& \int_{-\infty}^{+\infty} \int_{-\infty}^{+\infty}  \widehat{f}\widehat{f}^{*}e^{-i(\omega-\omega') t} d \omega d \omega' dt \nn \\
&=&\frac{1}{2\pi} \int_{-\infty}^{+\infty} \widehat{f}\widehat{f}^{*} d \omega \nn \\
&=&\int_{0}^{+\infty} \frac{|\widehat{f}|^2}{\pi} d\omega.
\ea

Some well known formulae related to the hyperbolic trigonometric and Hankel functions are:
\ba
H_\nu^{(1)}(z)&=&\frac{1}{i \pi}\int_{-\infty}^{+\infty} e^{z \sinh(t)-\nu t}dt,\\
\sinh(n x)&=&\frac12 (e^{n x}-e^{-n x}),\\
\cosh(n x)&=&\frac12 (e^{n x}+e^{-n x})
\ea
and with them we can easily calculate the Fourier transform of $\sinh$ as follows
\ba
\widehat{\sinh(n \xi)}&=&\int_{-\infty}^{+\infty} \sinh(n \xi) e^{i \omega t(\xi)}dt \nn \\
&=& \int_{-\infty}^{+\infty} \sinh(n \xi) e^{i \omega t(\xi)}\frac{dt}{d\xi} d\xi \nn \\
&=& \int_{-\infty}^{+\infty} \sinh(n \xi)\frac{1}{i \omega} \frac{d}{d \xi} e^{i \omega t(\xi)} d\xi \nn \\
&=&  \frac{1}{\pi i} \frac{\pi}{\omega} \left(\underbrace{(\cdots)}_{\textrm{equal to 0}}-n\int_{-\infty}^{+\infty} \cosh(n \xi) e^{i \omega t(\xi)} d\xi \right) \nn \\
&=& -\frac{n\pi}{2\omega} \frac{1}{\pi i} \int_{-\infty}^{+\infty} (e^{n \xi}+e^{-n \xi}) e^{i \omega t(\xi)} d\xi  \nn \\
&=& -\frac{n\pi}{2\omega} \frac{1}{\pi i} \int_{-\infty}^{+\infty} \left(e^{i \omega \nu_0 (e \sinh(\xi)-\xi)+n \xi}\right.\nn\\&+&\left.e^{i \omega \nu_0 (e \sinh(\xi)-\xi)-n \xi}\right) d\xi \nn \\
&=& -\frac{n\pi}{2\omega} \frac{1}{\pi i} \int_{-\infty}^{+\infty} \left(e^{i \omega \nu_0 e \sinh(\xi)- (i\omega \nu_0-n)\xi}\right.\nn\\&+&\left.e^{i \omega \nu_0 e \sinh(\xi)-(i\omega \nu_0+n) \xi}\right) d\xi \nn \\
&=& -\frac{n\pi}{2\omega} \frac{1}{\pi i} \int_{-\infty}^{+\infty} \left(e^{i \nu e \sinh(\xi)- (iv-n)\xi}\right.\nn\\&+&\left.e^{i \nu e \sinh(\xi)-(iv+n) \xi}\right) d\xi \nn \\
&=& -\frac{n\pi}{2\omega}\left(H_{i \nu-n}^{(1)}(i \nu e)+H_{i \nu+n}^{(1)}(i \nu e)\right),
\ea
where we have set $\nu=\nu_0 \omega$. Similarly, one can show that
\ba
\widehat{\cosh(n \xi)}&=&\int_{-\infty}^{+\infty} \cosh(n \xi) e^{i \omega t(\xi)}dt  \nn \\
&=& -\frac{n\pi}{2\omega}\left(H_{i \nu-n}^{(1)}(i \nu e)-H_{i \nu+n}^{(1)}(i \nu e)\right).~
\ea

Therefore, we have that
\ba
\widehat{\sinh(\xi)} &=& -\frac{\pi}{2\omega}\left(H_{i \nu-1}^{(1)}(i \nu e)+H_{i \nu+1}^{(1)}(i \nu e)\right) \\
\widehat{\cosh(\xi)} &=& -\frac{\pi}{2\omega}\left(H_{i \nu-1}^{(1)}(i \nu e)-H_{i \nu+1}^{(1)}(i \nu e)\right) \\
\widehat{\sinh(2\xi)} &=& -\frac{\pi}{\omega}\left(H_{i \nu-2}^{(1)}(i \nu e)+H_{i \nu+2}^{(1)}(i \nu e)\right) \\
\widehat{\cosh(2\xi)} &=& -\frac{\pi}{\omega}\left(H_{i \nu-2}^{(1)}(i \nu e)-H_{i \nu+2}^{(1)}(i \nu e)\right)
\ea
Using the identities
\ba
H_{\alpha}^{(1)}(x)&=&\frac12\frac{x}{\alpha}\left(H_{\alpha-1}^{(1)}(x)+H_{\alpha+1}^{(1)}(x)\right),\\
H_{\alpha}^{(1)}{}'(x)&=&\frac12\left(H_{\alpha-1}^{(1)}(x)-H_{\alpha+1}^{(1)}(x)\right)
\ea
and playing with the algebra we can show the following identities for the Hankel functions:
\ba
H_{\alpha-1}^{(1)}(x)+H_{\alpha+1}^{(1)}(x) &=& \frac{2\alpha}{x} H_{\alpha}^{(1)}(x), \\
H_{\alpha-1}^{(1)}(x)-H_{\alpha+1}^{(1)}(x) &=& 2H_{\alpha}^{(1)}{}'(x),\\
H_{\alpha-2}^{(1)}(x)+H_{\alpha+2}^{(1)}(x) &=& 2H_{\alpha}^{(1)}{}''(x)-\frac{2}{x}H_{\alpha}^{(1)}{}'(x)+\frac{2\alpha^2}{x^2}H_{\alpha}^{(1)}(x)\nn\\
&=& \left(\frac{4\alpha^2}{x^2}-2\right)H_{\alpha}^{(1)}(x)-\frac{4}{x}H_{\alpha}^{(1)}{}'(x),\nn \\~&&\\
H_{\alpha-2}^{(1)}(x)-H_{\alpha+2}^{(1)}(x) &=& \frac{4\alpha}{x}H_{\alpha}^{(1)}{}'(x)-\frac{4\alpha}{x^2}H_{\alpha}^{(1)}(x).
\ea

In terms of the dimensionless frequency $\nu$ and the eccentricity $e$ the above can be written as
\begin{widetext}
\ba
H_{i \nu-1}^{(1)}(i \nu e)+H_{i \nu +1}^{(1)}(i \nu e) &=& \frac{2}{e} H_{i \nu}^{(1)}(i \nu e) \\
H_{i \nu-1}^{(1)}(i \nu e)-H_{i \nu +1}^{(1)}(i \nu e) &=& 2H_{i \nu}^{(1)}{}'(i \nu e)\\
H_{i \nu-2}^{(1)}(i \nu e)+H_{i \nu +2}^{(1)}(i \nu e) &=& 2H_{i \nu}^{(1)}{}''(i \nu e)-\frac{2}{i \nu e}H_{i \nu}^{(1)}{}'(i \nu e)+\frac{2}{e^2}H_{i \nu}^{(1)}(i \nu e)\nn\\
&=& \left(\frac{4}{e^2}-2\right)H_{i \nu}^{(1)}(i \nu e)+\frac{4i}{\nu e}H_{i \nu}^{(1)}{}'(i \nu e)\\
H_{i \nu-2}^{(1)}(i \nu e)-H_{i \nu+2}^{(1)}(x) &=& \frac{4}{e}H_{i \nu}^{(1)}{}'(i \nu e)+\frac{4i}{ve^2}H_{i \nu e}^{(1)}(i \nu e)
\ea
\end{widetext}

and combining the above we finally get,
\ba
\widehat{\sinh(\xi)} &=& -\frac{\pi}{2\omega}\left( \frac{2}{e} H_{i \nu}^{(1)}(i \nu e) \right) \\
\widehat{\cosh(\xi)} &=& -\frac{\pi}{2\omega}\left( 2H_{i \nu}^{(1)}{}'(i \nu e) \right)
\ea

\ba
\widehat{\sinh(2\xi)} &=& -\frac{\pi}{\omega}\left( \left(\frac{4}{e^2}-2\right)H_{i \nu}^{(1)}(i \nu e)+\frac{4 i}{ \nu e}H_{i \nu}^{(1)}{}'(i \nu e) \right)\nn\\~ \\
\widehat{\cosh(2\xi)} &=& -\frac{\pi}{\omega}\left( \frac{4}{e}H_{i \nu}^{(1)}{}'(i \nu e)+\frac{4 i}{ ve^2}H_{i \nu e}^{(1)}(i \nu e)\right)
\ea

Using the formulae 9.3.15-16 and 9.3.19-20 of Abramovitz and Stegun \cite{Abramowitz:1974}, we also find the following very useful approximations:
\ba
H_{i \nu}^{(1)}(i \nu e)&\simeq& -i~\sqrt{\frac{2}{\pi \nu}}~\frac{1}{\sqrt[4]{e ^2-1}}~e^{\nu \left(\sec ^{-1}(e )-\sqrt{e ^2-1}\right)}\nn\\&\cdot&\left(1-\frac{M_1}{\nu}+\frac{L_2}{\nu^2}\right),\\
H_{i \nu}^{(1)}{}'(i \nu e)&\simeq& \sqrt{\frac{2}{\pi \nu}}~\frac{ \sqrt[4]{e ^2-1}}{e }~e^{\nu \left(\sec ^{-1}(e )-\sqrt{e ^2-1}\right)}\nn\\&\cdot&\left(1+\frac{O_1}{\nu}-\frac{N_2}{\nu^2}\right),
\ea
where we have set
\ba
M_1&=& \frac{5 \cot ^3(\beta )+3 \cot (\beta) }{24},\\
L_2&=& \frac{385 \cot ^6(\beta )+462 \cot ^4(\beta )+81 \cot ^2(\beta )}{1152},~~~~~~~~~\\
O_1&=& \frac{7 \cot ^3(\beta )+9 \cot (\beta )}{24}, \\
N_2&=& \frac{455 \cot ^6(\beta )+594 \cot ^4(\beta )+135 \cot ^2(\beta )}{1152}, \\
\cot (\beta )&=&\frac{1}{\sqrt{e ^2-1}}.
\ea

\bibliography{GWbursts}

\end{document}